# Spin-1 Dirac half-metal, spin-gapless semiconductor, and spin-polarized massive Dirac dispersion in transition metal dihalide monolayers

Muhammad Nadeem[†], Xiaolin Wang

*Institute for Superconducting and Electronic Materials, Australian institute for innovative materials, and ARC Centre of Excellence in Future Low-Energy Electronics Technologies (FLEET), University of Wollongong, Australia, NSW 2500, Australia.*

Spin-1 condensed matter systems characterized by the combination of a Dirac-like dispersion and flat bands are ideal for realizing high-temperature electronics and spintronic technologies in the absence of external magnetic field. In this study, we propose a three-band tight binding model, with spin-polarized Haldane-like next-nearest-neighbour tunnelling, on dice lattice and show that spin-1 Dirac half-metal, spin-1 Dirac spin-gapless semiconductor, and spin-polarized spin-1 massive Dirac dispersion with nontrivial topology can exist in two-dimensional ferromagnetic condensed matter systems with electron spin polarization $P = 1$. The proposed spin-polarized spin-1 phases can be realized in ferromagnetic transition metal dihalides $MX_2$ monolayers effectively. By using first principle calculations, we show that a small compressive strain leads $MX_2$ monolayers to be spin-one Dirac half-metal for $M = Fe$ and $X = Br, Cl$ while spin-one Dirac spin-gapless semiconductor for $M = Co$ and $X = Br, Cl$. Spin-one Dirac spin-gapless semiconductors $CoBr_2$ and $CoCl_2$ embeds flat band ferromagnetism where spin-orbit coupling opens a topologically non-trivial Dirac gap between dispersing valance and conduction band while leaving flat band unaffected. The intrinsic flat-band ferromagnetism in spin-polarized spin-1 massive Dirac dispersion plays key role in materializing quantum anomalous Hall state with Chern number $C = -2$.

## I. INTRODUCTION

Discovery of exotic topological phases hosting novel low energy quasiparticles is one of the key research areas in condensed matter physics. In recent years, Weyl[1-5] semimetals, Dirac[6,7] semimetals, and superconducting heterostructures[8-11] have been discovered where Dirac, Weyl, and Majorana fermions exist as low-energy excitations in these condensed matter systems[12-14]. Moreover, unlike high energy physics where the constraint of Lorentz invariance classifies the fermions in spin S=1/2 Dirac, Weyl, and Majorana particles only, spin S=1 and S=3/2 low energy fermion-like excitations are also allowed in condensed matter systems[15-23] where unconventional three, four, six and eight-fold band crossings are protected by space group symmetries[15]. Such condensed matter systems with multi-band crossings have unique surface states and transport properties[17].

Among these exotic three dimensional topological phases, condensed matter systems with S=1 chiral fermions characterized by the combination of a Dirac-like dispersion and flat bands[18-23] are ideal for realizing high-temperature electronics and spintronic technologies in the absence of external magnetic field. This unique feature of S=1 condensed matter systems is indebted to the presence of following two key features: (i) topologically trivial flat bands and (ii) the nontrivial overall bulk band topology. These features have also been studied in two dimensional pseduspin-1 optical lattices[24-26]. While Lieb lattice[25,26] incorporates both of these features, a simple dice lattice[24] lacks the later one.

However, if the bulk band dispersion can be made topologically nontrivial, dice lattice would have advantage of having large topological charge than Lieb lattice. This problem can be addressed for dice lattice in half-metallic or spin-gapless ferromagnetic condensed systems with electron spin polarization $P = 1$, unlike regular ferromagnetic materials where $0 < P < 1$. Keeping this microscopic feature of intrinsic magnetization in view, spin-polarized S=1 Dirac dispersion in dice lattice can be modelled through a three-band tight binding model (1). When only nearest-neighbour hoping is considered, single-particle band structure represents either a S=1 Dirac spin-gapless semiconductor or S=1 Dirac half-metal depending upon the on-site energies. When spin-polarized Haldane-like next-nearest-neighbour tunnelling[27] is also considered, flat band at the Fermi level remains unaffected while a topological gap opens between dispersing valance and conduction bands. As a result, a spin-polarized S=1 massive Dirac dispersion gives topologically non-trivial bulk band dispersion with Chern number -2.

Because of the presence of flat band, we refer these S=1 phases as flat band Dirac half-metals (FB-DHM) and flat band spin-gapless semiconductor (FB-SGS). While Dirac half-metal (DHM)[28] and spin-gapless semiconductors (SGS)[29] host spin-polarized S=1/2 Dirac dispersion, FB-DHM and FB-SGS hosts spin-polarized S=1 Dirac dispersion where a spin-polarized S=1/2 Dirac dispersion is accompanied by a single spin flat band. FB-SGS, where spin polarized flat band is lying exactly at the Fermi level, provides an ideal platform for the generalization of following two very important concepts previously studied in S=1/2 condensed matter systems – intrinsic ferromagnetism in two-dimensional materials[30-33] and quantum anomalous Hall (QAH) effect[27] in SGS[34-36] – to S=1 condensed matter systems.

In realistic condensed matter systems, transition metal dihalides $MX_2$ monolayers, where transition metal cation M = V, Cr, Mn, Fe, Co, Ni is surrounded by six halogen anions X=Br, Cl, I in octahedral environment, are much simpler two-dimensional systems that can host FB-SGS, FB-DHM and spin-polarized S=1 massive Dirac dispersion. It has been repotred[37,38] that $VX_2$, $CrX_2$, and $MnX_2$ monolayers are antiferromagnetic while $FeX_2$, $NiX_2$, $CoBr_2$ and $CoCl_2$ monolayers are ferromagnetic. Moreover, spin-resolved density of states shows that $M^{+2}$ ($Fe^{+2}(3d^6)$, $Co^{+2}(3d^7)$, $Ni^{+2}(3d^8)$) ions in $MX_2$ monolayers exhibit high-spin configuration: one of the spin sectors (say spin up) is distributed over whole $d$-manifold while the other spin sector (spin-down) fills the low-lying $t_{2g}$ orbitals with one, two, and three electrons in $FeX_2$, $CoX_2$, and $NiX_2$ monolayers respectively. Such a spin configuration of $d$-manifold renders $FeX_2$ monolayers as conventional ferromagnetic half-metals[38,39], $CoBr_2$ and $CoCl_2$ monolayers as ferromagnetic semiconductors with a very small band gap[40], while $NiX_2$ monolayers as ferromagnetic insulators[40].

$FeX_2$ and $CoX_2$ monolayers are of particular interest among these $MX_2$ family members for S=1 physics, mainly because of partially filled low-lying $t_{2g}$ orbitals. Here, by using first principle calculations, we show that a small compressive strain leads $FeBr_2$ and $FeCl_2$ monolayers to FB-DHM while $CoBr_2$ and $CoCl_2$ monolayers to FB-SGS. We further investigate the intrinsic ferromagnetic and topological bulk properties in $CoBr_2$ and $CoCl_2$ monolayers and found that: (i) the ferromagnetic ground state of $CoBr_2$ and $CoCl_2$ monolayers embeds flat band ferromagnetism[41-44]. (ii) spin-orbit coupling opens a topologically non-trivial Dirac gap between dispersing valance and conduction band while leaving flat band unaffected. This is in complete agreement with the effects of Haldane-like NNN tunnelling on the single-particle band structure obtained from three-band tight binding model on dice lattice. The combined effect of intrinsic flat-band ferromagnetism and nontrivial bulk electronic topological properties leads to quantum anomalous Hall state with Chern number $C = -2$.

## II. THREE-BAND TIGHT BINDING MODEL

The dice lattice[24] structure, with three-sites per unit cell, is composed of two inter penetrating honeycomb lattices, figure (1). While honeycomb lattice hosts S=1/2 Dirac dispersion, dice lattice is featured by S=1 Dirac dispersion. With ferromagnetic ground state, coupling between itinerant electron's spin and magnetization leads the dice lattice to host spin-polarized S=1 Dirac dispersion. FB-DHM, FB-SGS and spin-polarized S=1 massive Dirac band dispersion can be modelled through three-band tight binding Hamiltonian on the dice lattice

$$H = (\varepsilon_0 - M)\sum_i c_i^\dagger c_i - t\sum_{\langle ij \rangle} c_i^\dagger c_j - i\lambda \sum_{\langle\langle ij \rangle\rangle} v_{ij} c_i^\dagger c_j + m\sum_i v_i c_i^\dagger c_i \qquad (1)$$

Here $c_{i\alpha}^{\dagger}(c_{i\alpha})$ is the creation (annihilation) electron operator on site $i$, $\nu_i = (\nu_a, \nu_c, \nu_b) = (1, 0, -1)$ represents the sign convention for staggered sublattice potential, and $\nu_{ij} = \mathbf{d}_{ik} \times \mathbf{d}_{kj} = \pm 1$ connects sites $i$ and $j$ on sublattice A (and B) via the unique intermediate site $k$ on sublattice C. Here $\mathbf{d}_{ik}$ and $\mathbf{d}_{kj}$ are the nearest-neighbour bond vectors connecting A (and B) sublattices with sublattice C. In the model Hamiltonian (1), first term incorporates both on-site energy $\varepsilon_0$ and Zeeman exchange interaction $M > 0$, second term is the nearest neighbour hoping while third term is modified Kane and Mele type SO coupling[45,46] for S=1 system. By setting electron spin polarization $s_z = -1$, Kane and Mele type SO coupling term effectively reduces to Haldane-like term for next-nearest-neighbour tunnelling[27]. The last term represents the staggered sublattice potential. In the long wavelength limit, effective Hamiltonian and low energy single-particle band dispersion in the vicinity of spin-polarized S=1 Dirac points are

$$H(\pm K + q) = \begin{pmatrix} \varepsilon_0 - M + m \mp 3\sqrt{3}\lambda & \pm 3atq_{\pm}/2 & 0 \\ \pm 3atq_{\mp}/2 & \varepsilon_0 - M & \pm 3atq_{\pm}/2 \\ 0 & \pm 3atq_{\mp}/2 & \varepsilon_0 - M - m \pm 3\sqrt{3}\lambda \end{pmatrix} \quad (2)$$

$$E_0(\eta K + q) = \varepsilon_0 - M \quad (3a)$$

$$E_{\pm}(\eta K + q) = \varepsilon_0 - M \pm \sqrt{v_F^2 |q|^2 + 2|m - 3\sqrt{3}\eta\lambda|} \quad (3b)$$

where $v_F = 3at/\sqrt{2}$ is Fermi velocity, $q_{\pm} = q_x \pm q_y$ and $\eta = +1(-1)$ is the valley index representing $K(K')$. In the absence of SO coupling and staggered sublattice potential, three-fold degenerate crossing at K and K' represents S=1 spin-polarized Dirac points. That is, the Hamiltonian (1) represent FB-SGS and FB-DHM for $(\varepsilon_0 - M) = 0$ and $(\varepsilon_0 - M) > 0$ respectively, as shown in figure (1). When SO coupling and sublattice potential are activated, a Dirac gap of $E_g(\eta) = 2|m - 3\sqrt{3}\eta\lambda|$ is opened between dispersing valance and conduction bands in FB-SGS. As shown in figure (2), spin-polarized S=1 massive Dirac dispersion captures the salient features of Haldane model for S=1/2 systems: the gap remains always positive at valley K', however, it decreases at valley K with increasing the strength of staggered potential $m$, closes when $m = 3\sqrt{3}\lambda$, and reopens for $m > 3\sqrt{3}\lambda$. Band dispersion of semi-infinite nanoribbon confirms the non-zero Chern number, $C = -2$. In next section, we show that electronic band structure obtained from three-band tight-binding model for infinite dice sheet qualitatively reproduces the electronic properties of strained $CoBr_2$ and $CoCl_2$ monolayer predicted by first principle calculations and hence reveal the underlying topological physics effectively.

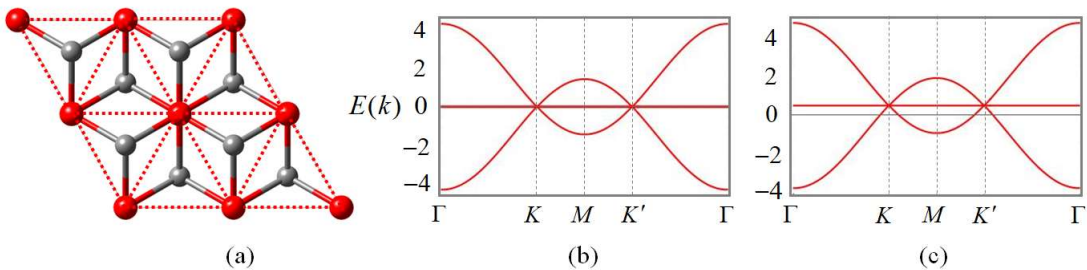

(a)            (b)            (c)

**Figure 1: Lattice and single-particle band dispersion for three-band tight binding model.** A two-dimensional network of two intermingled honeycomb lattices, known as dice structure with three-sites per unit cell, hosting S=1 Dirac dispersion **(a)**. When $\lambda = m = 0$, three-band tight binding model (1) gives FB-SGS for $\varepsilon_0 - M = 0$ **(b)** and FB-DHM for $\varepsilon_0 - M > 0$ **(c)**.

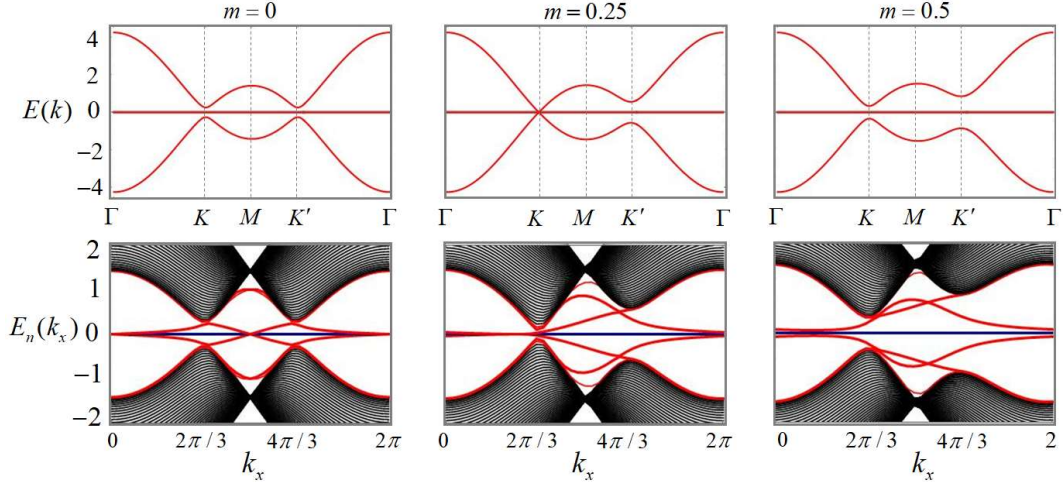

**Figure 2: Effects of SO coupling and staggered sublattice potential on FB-SGS.** (*Upper panel*) Bulk band structure for infinite sheet of dice lattice, showing topological phase transition from QAH phase to trivial insulating phase. (*Lower panel*) Band structure for semi-infinite sheet with zigzag edges. For $0 \leq m < 3\sqrt{3}\lambda$, edge state crossing Fermi level shows nontrivial behaviour, $C = -2$. Here we set $\lambda = 0.05$ and $\varepsilon_0 - M = 0$.

## III. MATERIAL REALIZATION

It is worthy to shed some light on microscopic ferromagnetic phenomena for bridging between three-band tight binding model on dice lattice and first principle caculations for $MX_2$ monolayers presented in this section. In a conventional ferromagnetic system, where electron spin polarization is larger than 0 but smaller than 1; $0 < P < 1$, a realistic Zeeman-type exchange splitting could not completely distangle the two oppositely spin polarized bands near Fermi level. That is, if we include both spin sectors in Hamiltonian (1), S=1 spin-up(spin-down) Dirac points will be lying $\varepsilon_0 + M$ above ($\varepsilon_0 - M$ below) the Fermi level and there would be band crossing between oppositely spin polarized bands at Fermi level which is located at the energy zero. As a result, when Zeeman coupling $M$ is less than the bandwidth $W$ of topologically non-trivial dispersing bands, even a spin-orbit coupled ferromagnetic insulator would turn into a ferromagnetic metal – a partial concern for the realization of QAH phase in conventional ferromagnetic systems.

This problem can be naturally addressed in ferromagnetic spin-gapless semiconductors, ferromagnetic half-metals, or spin-orbit coupled ferromagnetic insulators where electron spin polarization is $P = 1$; oppositely spin-polarized sectors are already well separated in energy and only one of the spin-polarized bands reside at/near Fermi level. With ferromagnetic ground state, by setting all localized Ising spins directed out-of-plane (along z-axis) and considering itinerant electrons carry spin-down, coupling between itinerant electron's spin and magnetization provided by localized spins leads spin-polarized band dispersion.

Secondly, Hamiltonian (1) effectively describes the FB-DHM, FB-SGS and spin-polarized S=1 massive Dirac band dispersion with characteristics similar as we find in first principle calculations for transition metal dihalide monolayers. However, in realistic transition metal dihalides, the edges of conduction and valence bands near Fermi level are predominantly contributed by all three low-lying $t_{2g}$ orbitals of transition metal ions. Hence, a more accurate tight binding model should be constructed based on all three orbitals along with the consideration of symmetries of the monolayers. Moreover, for $3d$ orbitals involving multiple bands, possible electron-electron interactions are required to be considered for complete and quantitative understanding of flat band magnetism.

## A. First principle calculations

The presence of FBSGS, FB-DHM, and spin-polarized S=1 massive Dirac dispersion with nontrivial bulk topology requires large intrinsic magnetization and strong SO coupling. We show that these features are embedded in transition metal dihalides $MX_2$ monolayers. We study the magnetic and electronic properties of $MX_2$ (M = Fe, Co; X=Br, Cl) monolayers with space group P-3m1 using the projected augmented plane-wave method (PAW)[47,48] as implemented in the Vienna ab initio simulation package (VASP)[49]. The Perdew-Burke-Ernzerhof functional of the generalized gradient approximation (GGA) is employed for the account of exchange-correlation interactions among electrons[50]. The cutoff energy for the plane-wave-basis set is chosen to be 400 eV for all calculations. A 2×2×1 supercell is used to calculate the exchange energy of ferromagnetic and antiferromagnetic (with periodic boundary conditions) configurations of Co spins. A 35×35×1 k-point grid is employed to sample the Brillouin zone. The structure is obtained by fully relaxing the atomic positions until the Hellmann-Feynman forces on each ion are less than 0.01 eV/A$^{-1}$. Convergence of total energies and full optimization of the atomic coordinates yields the hexagonal lattice constants for $FeX_2$ and $CoX_2$ as shown in table 1. These values are in close agreement as previously reported[38,40]. To study the massive Dirac dispersion, SO coupling is included in the self-consistent loop.

| $MX_2$ | Pristine | | Strained | |
|---|---|---|---|---|
| $FeBr_2$ | 3.725 | HM | 3.56 | FB-DHM |
| $FeCl_2$ | 3.526 | HM | 3.35 | FB-DHM |
| $FeI_2$ | 4.07 | HM | - | - |
| $CoBr_2$ | 3.768 | SC | 3.70 | FB-SGS |
| $CoCl_2$ | 3.51 | SC | 3.47 | FB-SGS |

**Table 1:** Calculated lattice parameter (a = b, in Å) for pristine and strained $MX_2$ monolayers and the corresponding ferromagnetic electronic phases.

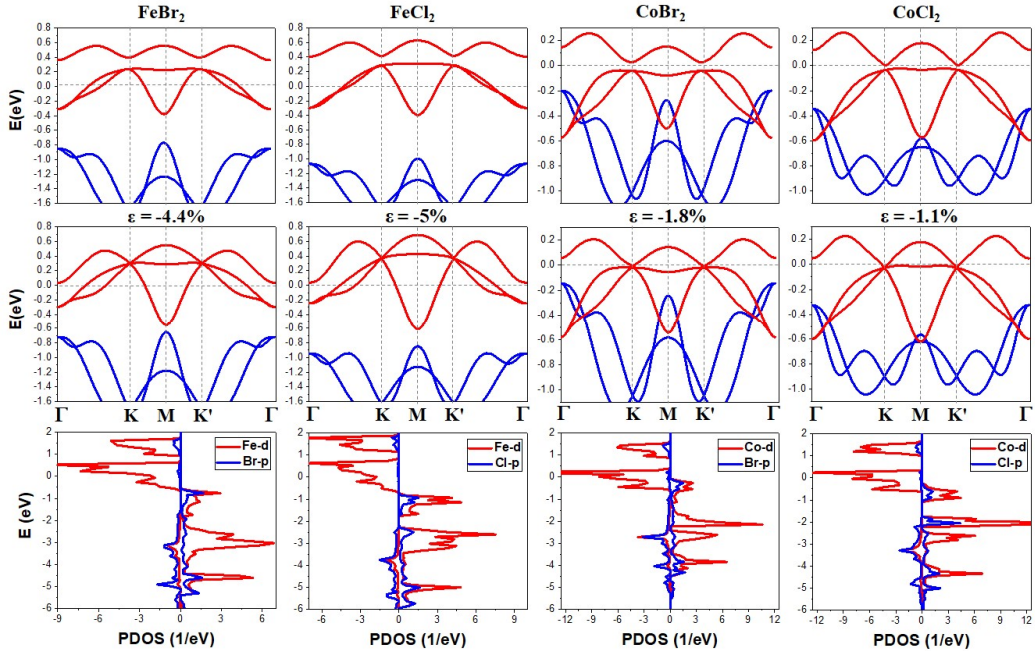

**Figure 3: First principle calculation for $FeBr_2$, $FeCl_2$, $CoBr_2$, and $CoCl_2$ monolayers.** Red (blue) curves represent spin-down(spin-up) bands. Top and middle panels show band dispersion for pristine and strained $FeBr_2$, $FeCl_2$, $CoBr_2$, and $CoCl_2$ monolayers respectively. A small compressive strain leads $FeBr_2$ and $FeCl_2$ monolayers to S=1 Dirac half-metallic phase while $CoBr_2$ and $CoCl_2$ monolayers to S=1 spin-polarized Dirac gapless state. Lower panel shows partial density of states (PDOS) for strained $FeBr_2$, $FeCl_2$, $CoBr_2$, and $CoCl_2$ monolayers.

## B. FB-SGS and FB-DHM

In terms of electronic structure, consistent with the previous results[38,39], first principle calculations show that $FeX_2$ monolayers are conventional ferromagnetic half-metals while $CoBr_2$ and $CoCl_2$ monolayers are ferromagnetic semiconductors with a very small band gap[40]. However, a compressive strain leads $FeBr_2$ and $FeCl_2$ monolayers to FB-DHM while $CoBr_2$ and $CoCl_2$ monolayers to FB-SGS. As shown in figure (3), the band at the Fermi level remains flat near the corners of Brillion zone (high symmetry points K and K') for $CoBr_2$ monolayer while the band at the Fermi level remains flat along the boundary of Brillion zone (high symmetry line K-M-K') for $CoCl_2$ monolayer. Like $FeBr_2$ and $FeCl_2$, pristine $FeI_2$ monolayer is also a conventional ferromagnetic half metal. However, a very large compressive strain is required to attain FB-DHM in $FeI_2$ monolayer. Spin-polarized S=1 Dirac band dispersions for FB-DHM and FB-SGS can easily be attained from electronic band dispersion of pristine monolayers under very small and experimentally accessible compressive strains of $\varepsilon = -4.4\%$, $\varepsilon = -5\%$, $\varepsilon = -1.8\%$, and $\varepsilon = -1.1\%$ for $FeBr_2$, $FeCl_2$, $CoBr_2$ and $CoCl_2$ monolayers respectively.

## C. Flat band ferromagnetism

We calculate the ground state energy for both ferromagnetic and periodic antiferromagnetic (stripe order) spin configuration for $M_4X_8$ cell as shown in figure (4). The large positive values for exchange energy confirm that the ground state of $FeBr_2$, $FeCl_2$, $CoBr_2$ and $CoCl_2$ monolayers are ferromagnetic with a high Curie temperature. Ferromagnetic ordering is consistent with Goodenough–Kanamori–Anderson (GKA) rules[51-53]. While the localized magnetic moments arise through the competition between ionic and covalent bond strength between M and X ions, the magnetic ordering originates from the interplay between superexchange and direct exchange interaction. Although both direct and superexchange leads to antiferromagnetic ordering in principle, GKA rules indicate that the dominating superexchange interaction need not to favour antiferromagnetic ordering in all cases: If $d$-orbitals on neighbouring M ions hybridize with orthogonal $p$ orbitals on X ions such that M-X-M bond angles are $90^0$, the superexchange gives ferromagnetic ordering.

Spin-resolved density of states shown in figure (3) reflect two key features of $FeBr_2$, $FeCl_2$, $CoBr_2$ and $CoCl_2$ monolayers: (i) With seven (six) $3d$ electrons, $Co^{+2}$ ($Fe^{+2}$) ions exhibit high-spin configuration, five with spin-up and two (one) electrons with spin-down state. The calculated total magnetic moment of $3\mu_B$ ($4\mu_B$) per unit cell, with most of it carrying by Co ≈ 2.48 (Fe ≈ 3.45) ions, confirms that the ground state favours high spin-configuration. In the octahedral crystal field environment, where five $3d$ orbitals split into two $e_g$ and three $t_{2g}$ orbitals, five spin up electrons are distributed over whole $3d$ manifold while two (one) spin-down electrons reside in low energy $t_{2g}$ orbitals. This spin configuration of $Co^{+2}$ ($Fe^{+2}$) electrons leaves $CoBr_2$ and $CoCl_2$ ($FeCl_2$ and $FeBr_2$) monolayers in a spin-polarized gapless (half metal) state where only spin-down electrons in $t_{2g}$ orbitals reside near the Fermi energy.

(ii) A sharp peak of density of states at Fermi level indicates the presence of spin-polarized flat band and hence flat-band ferromagnetism in $CoBr_2$ and $CoCl_2$ monolayers. This prominent feature is the result of specific inter-site electron hoping in $MX_2$ monolayers which is composed of two intermingled honeycomb lattices with unequal number of M and X sites[41]. The effect of on-site correlations on magnetic ordering and spin-configuration of $Co^{+2}$ ($Fe^{+2}$) ions is studied through LDA+$U$. The flat band ferromagnetism[41-44], for any positive value of on-site Coulomb correlations $U \geq 0$, in $CoBr_2$ and $CoCl_2$ monolayers can be justified in the non-interacting or weakly-interacting limit: when flat band at the Fermi level is saturated with fully spin-polarized electrons, the Pauli exclusion of double occupancy minimizes the Coulomb interaction and renders ground state of $CoBr_2$ and $CoCl_2$ monolayer ferromagnetic.

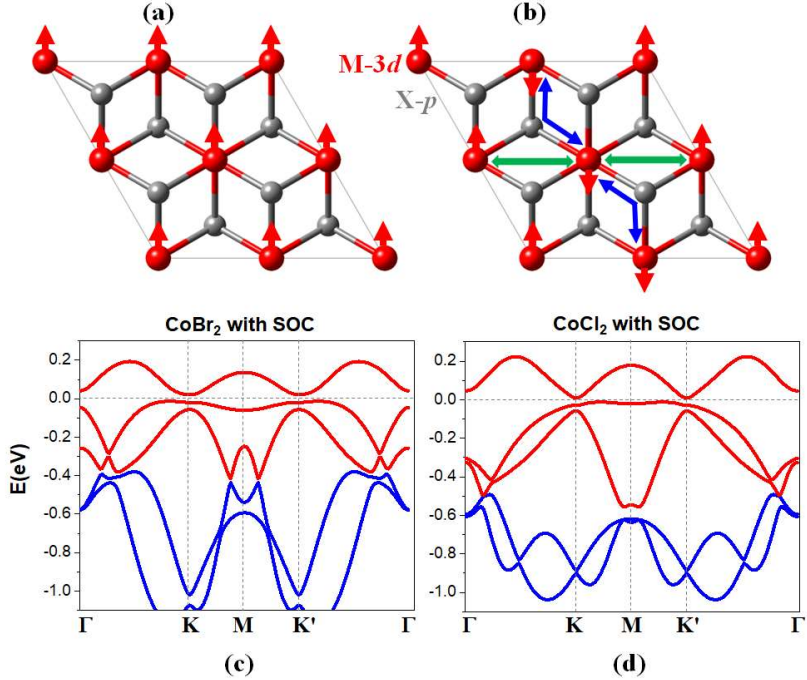

**Figure 4: Magnetic configuration and effects of SO coupling in $CoBr_2$ and $CoCl_2$ monolayers.** **(a, b)** Ferromagnetic **(a)** and periodic antiferromagnetic **(b)** spin configuration for $MX_2$. Blue and green arrows represent the mechanism for ferromagnetic superexchange interaction with $90^0$ of M-X-M bond angle and antiferromagnetic direct M-M exchange interaction between localized moments respectively. **(c, d)** Band structure of $CoBr_2$ **(c)** and $CoCl_2$ **(d)** monolayers with SO coupling. SO coupling open a gap between dispersing valance and conduction band in whole Brillion zone for both $CoBr_2$ and $CoCl_2$ monolayers.

### D. Spin-polarized S=1 massive Dirac dispersion - Effect of SO coupling

Having established spin-polarized S=1 Dirac character of $CoBr_2$ and $CoCl_2$ monolayers with intrinsic flat band ferromagnetism, we explore the possible effects of SO coupling on both Dirac dispersing and flat band as shown in figure (4). In the non-interacting limit, a gap is opened between dispersing valance and conduction bands by the activation of SO coupling. However, spin-polarized flat band lying at Fermi level remains unaffected. That is, flat band gets detached from both dispersing valance and conduction bands but remains at the Fermi level. This is in complete agreement with the single-particle band structure obtained from three-band tight binding model on dice lattice. Simultaneous presence of flat band ferromagnetism and topological spin-orbit coupling gap in FB-SGS is promising for the exploration of QAH phase when Fermi energy $E_F$ is positioned within the Dirac gap between dispersing bands.

Possibility of QAH phase, with a trademark of dissipationless chiral edge states, in two-dimensional S=1 spin-polarized Dirac systems has an obvious advantage over QAH phase in S=1/2 magnetic doped topological insulators (TIs)[54-56]. The idea of achieving ferromagnetic ordering in TIs by inducing magnetic impurities[54], has already led to the experimental realization of QAH phase in $Cr^{57,58}$ and $V^{59}$ doped thin films of tetradymite semiconductors $(Bi,Sb)_2Te_3$. However, the highest temperature at which QAH phase is observed in magnetic doped TIs is only few Kelvin. Several experimental challenges limit the high temperature realization of QAH phase in magnetic doped TIs. For example, presence of spin scattering centres due to non-uniform distribution of magnetic impurities and parallel dissipative channels at the one-dimensional edges are likely to induce several KΩ longitudinal resistance. Moreover, as doped magnetic impurities are required to generate a sufficiently strong magnetic moment without effecting the band topology of the system, difficulty in simultaneous achievement of high $T_c$ and a large topological gap opening in magnetically doped TIs limit the applications of high temperature QAH phase[33].

On the other hand, FB-SGS with intrinsic flat band ferromagnetism, sustainable against thermal fluctuations at high temperature in the two-dimensional limit, in combination with topologically non-trivial bulk band structure can offer QAH phase without requiring random magnetic impurity doping. With quantized transverse resistance, extremely low longitudinal resistance, and large topological charge could allow possible future dissipationless spintronics and low-energy electronics technologies in two-dimensional FB-SGS.

## IV. DISCUSSION

We proposed a three-band tight binding model, with spin-polarized Haldane-like next-nearest-neighbour tunnelling, for ferromagnetic condensed matter systems with electron spin polarization $P = 1$. By using first principle calculations, we further show that $FeBr_2$ and $FeCl_2$ monolayers are FB-DHM while $CoBr_2$ and $CoCl_2$ monolayers are FB-SGS with spin-polarized S=1 Dirac dispersion. When SO coupling is activated, like the effects of Haldane-like NNN tunnelling on the single-particle band structure obtained from three-band tight binding model on dice lattice, a topologically non-trivial Dirac gap is opened between dispersing valance and conduction band while leaving flat band unaffected.

FB-SGS embeds two interesting phenomena and has an obvious advantage over S=1/2 SGS; the presence of spin-polarized flat band at the Fermi level leads to two-dimensional intrinsic flat band ferromagnetism and topologically nontrivial electronic properties with large topological charge. Since the physical realization of quantum anomalous Hall effect is indebted to the existence of intrinsic ferromagnetism, this study greatly extends the search for material classes of quantum anomalous Hall phase from S=1/2 to S=1 condensed matter system. Moreover, in the presence of strong SO coupling, intrinsic flat band ferromagnetism can serve as guiding principle for the experimental realization of other time reversal symmetry breaking topological phases such as topological superconductors. Finally, intrinsic flat band ferromagnetism and topologically nontrivial electronic properties in two-dimensional S=1 condensed matter system is promising for playing vital role in the realization of dissipationless spintronics and electronics technologies.

**Acknowledgments**

We acknowledge the support from the Australian Research Council (ARC) through an ARC Discovery Project (DP130102956) and an ARC Professorial Future Fellowship project (FT130100778).

---

[†]*Author is on leave from Department of Basic Sciences, School of Electrical Engineering and Computer Science, National University of Sciences and Technology (NUST), H-12 Islamabad, Pakistan.*